\documentclass[preprint,aps]{revtex4-1}
\usepackage{dcolumn}
\usepackage{color}
\usepackage[normalem]{ulem}
\usepackage{amssymb,amsmath,amsthm,epsfig,subfigure} 
\usepackage{bm}
\def\beq{\begin{equation}}
\def\eeq{\end{equation}}
\def\beqar{\begin{eqnarray}}
\def\eeqar{\end{eqnarray}}
\newcommand{\ok}{\overline{k}}
\newcommand{\bx}{{\bf{x}}}

\newcommand{\tomega}{\tilde{\omega}}

\newcommand{\hf}{\hat{f}}

\newcommand{\U}{{\bf U}}

\newcommand{\bu}{{\bf u}}

\newcommand{\bk}{{\bf k}}

\newcommand{\p}{{\partial}}
\newcommand{\hh}{\tilde{h}}

\newcommand{\tv}{{\tilde { v}}}
\newcommand{\tf}{\tilde{f}}
\newcommand{\hv}{{\hat { v}}}

\newcommand{\bbeta}{{\tilde{\beta}}}
\newcommand{\gapprox}{\lower.4ex\hbox{$\;\buildrel
>\over{\scriptstyle\sim}\;$}}
\newcommand{\lapprox}{\lower.4ex\hbox{$\;\buildrel
<\over{\scriptstyle\sim}\;$}}

\begin{document}
\title{Effect of enhanced dissipation by shear flows on transient relaxation and probability density function in two dimensions}

\author{Eun-jin Kim and Ismail Movahedi}
\affiliation{
School of Mathematics and Statistics, University of Sheffield, Sheffield,
S3 7RH, U.K.}

\begin{abstract}
We report a non-perturbative study of the effects of shear flows on turbulence reduction in a decaying turbulence in two dimensions. By considering different initial
power spectra and shear flows (zonal flows, combined zonal flows and streamers), we demonstrate how shear flows 
rapidly generate small scales, leading to a fast damping of turbulence amplitude. In particular,
a double exponential decrease in turbulence amplitude is shown to occur due to an exponential increase in wavenumber. The scaling of the effective dissipation time scale $\tau_{e}$, previously taken to be a hybrid time scale $\tau_{e} \propto \tau_{\Omega}^{{2/3}} \tau_{\eta}$, is shown to depend on types of depend on the type of shear flow as well as the initial power spectrum. Here, $\tau_{\Omega}$ and $\tau_{\eta}$ are shearing and molecular diffusion times, respectively. Furthermore, we present time-dependent Probability Density Functions (PDFs) and discuss the effect of enhanced dissipation on PDFs and a dynamical time scale $\tau(t)$, which represents the time scale over which a system passes through statistically different states. 
\end{abstract}

\pacs{52.25.Fi, 52.35.Mw, 52.35.Ra, 52.55.Dy}
\maketitle

\vspace{0.1cm}
\section{Introduction}

Large scale shear flows are one of the most ubiquitous structures that naturally occur in a variety
of physical systems and play
an essential role in determining the overall transport in those systems.
For example, stable shear flows can dramatically quench
turbulent transport by shear-induced-enhanced-dissipation (see, e.g., 
\cite{BURRELL,HAHM02,DAM2017,Chang2017,KIM1,KIM2,KIM3,KIM5,KIM4,Diamond,LI,IDOMURA,LAO,SYNAKOWSKI,GARBET,REWOLDT}). This occurs as
a shear flow distorts fluid eddies, accelerates the formation of small scales, and dissipates
them when a molecular diffusion becomes effective on small scales.
One remarkable consequence of this turbulence quenching
is the formation of transport barrier where the transport is dramatically reduced. 
The transition from low-confinement to high-confinement mode (L-H transition) in laboratory plasmas results from
such formation of a transport barrier by shear flows (e.g. see \cite{BURRELL,HAHM02,KIM1,Diamond}), which is believed to be crucial for a successful operation of fusion devices. A similar transport barrier is also induced by a shear layer in
the oceans \cite{HUNT} and by an equatorial wind in the atmosphere \cite{ATM}.
In the solar interior, a prominent large-scale shear flow due to the radial differential
rotation was shown to lead to weak anisotropic turbulence and mixing in the tachocline \cite{KIM3,KIM5} -- the boundary layer between the stable radiative interior and unstable convective layer. Our theoretical predictions have been confirmed by various numerical simulations (e.g., see  \cite{aditi,Newton11}.)

The purpose of this paper is to investigate the effect of shear flows on the time-evolution of turbulence. In most of the previous works, the main focus was on the calculation of turbulent transport in a stationary state in a forced turbulence. Different models of turbulence such as 2D and 3D hydrodynamics and magnetohydrodynamic turbulence with/without rotation and stratification as well as different types of shear flows (e.g. linear, oscillatory, stochastic shear flows) \cite{KIM1,KIM2,KIM3,KIM5,KIM4} were considered previously. In comparison, much less work was done on the effect of shear flows on the dynamics/time-evolution of turbulence, more precisely, how the enhanced/accelerated dissipation is manifested in time-evolution. A clear manifestation of shear flow effects on the dynamics seems especially important given an ongoing controversy over the role of a shear flow in transport reduction, e.g., whether it is due to the reduction in cross phase (via an increased memory, as caused by waves) or the reduction in the amplitude of turbulence via enhanced dissipation (e.g. see \cite{Diamond, Newton11} and references therein). A decaying turbulence provides us with an excellent framework in which this can be investigated in depth. We thus consider a simple decaying two-dimensional hydrodynamic turbulence model and examine the transient relaxation of the vorticity by different types of shear flows. We present time-dependent Probability Density Functions (PDFs) and discuss the effects of enhanced dissipation by shear flows on PDFs and effective dissipation time scale $\tau_{e}$. We also introduce a dynamical time scale $\tau(t)$ which measures the rate of change in information associated with time-evolution; $1/\tau(t)$ represents the rate at which a system passes through statistically different states at time $t$ (see \S 3). 

The simplicity of our model permits us to perform detailed analysis for different power spectra and shear flows.
{ Nevertheless, our result that the dissipation and dynamical time scale depend on power spectrum and different types of shears 
is generic.}
The remainder of this paper is organised as follows. \S 2 introduces our model and highlights the importance of i) a careful treatment of
a diffusion term in a PDF method and ii) a non-perturbative treatment of shear flows. \S 3 introduces dynamical time unit $\tau(t)$.
\S 4 discusses the effect of different shear flows on the evolution of Gaussian PDFs for different power spectra. \S 5 presents the analysis of one example of non-Gaussian PDFs. Discussion and Conclusions are found in \S 6. Appendices contain some of the detailed mathematical derivations.

\section{Probability Density Function (PDF)}
We consider the evolution equation for the fluctuating vorticity $\omega$ in two dimensions (2D). In the presence of a large-scale shear flow ${\bf U}$, turbulence becomes weak \cite{KIM2,KIM3,KIM5,KIM4}, and we can thus consider 
the following linear equation for fluctuating vorticity $\omega$ ($= - \nabla^2 \phi$ where $\phi$ is a stream function, or electric
potential in plasmas)
\begin{eqnarray}
\left[ {\p_t}+ {\bf U} \cdot \nabla \right]\omega
 &=& {\nu} \nabla^2 \omega\,.
\label{eq3} 
\end{eqnarray}
For simplicity, Eq. (\ref{eq3}) is taken to be dimensionless after appropriate rescaling of $U$, $\omega$, $\nu$, $\bx$ and $t$. Note that since our main focus
is on elucidating the effect of shear flows, scaling relations and relative values are of interest.
Despite the fact that Eq. (\ref{eq3}) is linear in $\omega$, the equation for  $p(\omega,\bx,t)$ is not closed due to the dissipation term involving the second derivative. To show this, we express $p(\omega,\bx,t)$ as the Fourier transform of the average of a generating function $Z = \exp{(i \lambda \omega(\bx,t))}$ (e.g. see \cite{JUSTIN,POPE}) as 
\begin{equation}
p(\omega,\bx,t) = \langle \delta (\omega(\bx,t) - \omega) \rangle =\frac{1}{2\pi} \langle  \int d \lambda \,e^{-i \lambda (\omega- \omega(\bx,t))} \rangle =\frac{1}{2\pi} \int d \lambda\,  e^{-i \lambda \omega} \langle Z \rangle,
\label{eq4}
\end{equation}
where the angular brackets denote the average. By differentiating $Z$ and using Eq. (\ref{eq3}), we obtain
\begin{eqnarray}
{\p_t} Z 
&=& i \lambda (\p_t \omega) Z
=  i \lambda \left [- \U \cdot \nabla \omega + {\nu} \nabla^2 \omega \right] Z.
\label{eq5} 
\end{eqnarray}
By using $\p_j Z = i \lambda (\p_j \omega )Z$ and $\p_{jj} Z = i \lambda (\p_{jj} \omega) Z - \lambda^2 (\p_j \omega)^2 Z$,
we recast Eq. (\ref{eq5}):
 \begin{eqnarray}
{\p_t} Z + {\bf U} \cdot \nabla Z &=& \nu \left[ \p_{jj} Z -[\p_j(\ln{Z})^2] Z \right]
= \nu \left[ \nabla^{2} Z + \lambda^2 (\p_j \omega)^2 Z \right].
\label{eq6}
\end{eqnarray}
The second equation in Eq. (\ref{eq6}) shows that the diffusion term gives rise to a nonlinear term in $Z$ ($[\p_j (\ln{Z})^2] Z$). The Fourier transform of  $\langle [\p_j(\ln{Z})^2] Z \rangle$ would then induce a convolution of $p(\omega, \bx,t)$.  On the other hand, the Fourier transform of $\langle\lambda^2 (\p_j \omega)^2 Z \rangle$ in the last equation in Eq. (\ref{eq6}) would require a conditional probability \cite{POPE}. For statistically independent $\p_j \omega$ and $Z$, a linear equation can be written as
\begin{eqnarray}
{\p_t} p + {\bf U} \cdot \nabla p &=& \nu \nabla^2 p    - \nu \langle (\p_j\omega)^2 \rangle \p_{\omega \omega} p.
\label{eq7} 
\end{eqnarray}
For a homogeneous turbulence, $p(\omega,\bx,t)$ becomes independent of $\bx$, reducing Eq. (\ref{eq7}) to
\begin{eqnarray}
{\p_t} p  &=&    - \nu \langle (\p_j\omega)^2 \rangle \p_{\omega \omega} p.
\label{eq70} 
\end{eqnarray}
In general, the treatment of the diffusion term involving $\nu$ is tricky and has often been done approximately, or the diffusion term is simply neglected. Unfortunately, such an approximation cannot be justified in the presence of a shear flow as its effect is enhanced due to the accelerated formation of small scales, demanding the exact treatment of this diffusion term. For the same reason, the effect of $\U$ cannot be treated perturbatively. 

It is thus pivotal to solve Eq. (\ref{eq3}) exactly in the Fourier space by using a time-dependent wave number. For example, let us consider a general type of a shear flow ${\bf U} =(U_s, U_y) =  (-y\Omega_s, -x\Omega_z)$, where $U_z$ and $U_s$ are orthogonal flows, with their shearing rate $\Omega_z$ and $\Omega_s$, respectively. We call $U_{z}$ zonal flows and $U_{s}$ streamers in this paper. $\U$ has the mean vorticity $\langle \omega_{T}\rangle =  \nabla \times \U = (-\Omega_z + \Omega_s) {\hat z}$. In order to capture the effect of shear non-perturbatively, we use the following time-dependent wavenumber
(e.g. see \cite{KIM2,KIM3,KIM5,KIM4}):
\begin{equation}
\omega({{\bf x}},t) = {\tilde \omega}({\bf k},t) \exp{\{i(k_x(t) x + k_y (t)y)\}}\,,
\label{eq8}
\end{equation}
where $k_x(t)$ and $k_y(t)$ satisfy
\begin{eqnarray}
\frac{d k_x(t)}{dt} =  \Omega_z k_y\,\,\,\,\, \frac{d k_y(t)}{dt} = \Omega_z k_x.
\label{eq9}
\end{eqnarray}
Eqs. (\ref{eq8})-(\ref{eq9}) give us a linear equation for the Fourier component ${\tilde \omega}({\bf k},t)$ as
$\frac{\partial {\tilde \omega}({\bf k},t)}{\partial t} = - \nu [k_{x}(t)^{2}+k_{y}(t)^{2}]{\tilde \omega}({\bf k},t)$
with the solution
\begin{equation}
{\tilde \omega}({\bf k},t)= {\tilde \omega}({\bf k}(0),t=0) \exp{ \left( - \nu \int_{0}^{t}dt_{1}[ k_{x}(t_{1})^{2}+k_{y}(t_{1})^{2}]\right)}.
\label{eq91}
\end{equation}
Eq. (\ref{eq91}) would then permit us to compute $Z = \exp{(i \lambda \omega(\bx,t))}$ and thus $p(\omega,\bx,t)$ in Eq. (\ref{eq4}). Once we have $p(\omega,\bx,t)$, we can then find the equation for $p(\omega, \bx,t)$.

\section{Dynamical time unit $\tau(t)$}
Having introduced a time-dependent PDF in \S 2, we now present how to utilize it to extract useful information diagnostics.
A key characteristic of non-equilibrium processes is the variability in time (or in space), time-varying
PDFs manifesting the change in information content in the system. We quantify the change in information by the rate at which a system passes through statistically different states \cite{NK14,NK15,HK16,KIM16,KH17}. Mathematically, for a time-dependent PDF $p(x,t)$ for a stochastic variable $x$, we define  the characteristic timescale $\tau(t)$ over which $p(x,t)$ temporally changes {\it on average}
at time $t$ as follows:
\begin{eqnarray}
{\cal{E}} \equiv \frac{1}{[\tau (t)]^2} & = &\int dx \frac {1} {p(x,t)}
 \left [\frac {\partial p(x,t)} {\partial t} \right]^2.\label{eq01}
\end{eqnarray}
As
defined in Eq. (\ref{eq01}), $\tau(t)$ is a dynamical time unit, measuring the correlation time of $p(x,t)$.
Alternatively, $1/\tau$ quantifies the (average) rate of change of information in time.
A special case of $\tau(t)=$ constant is a geodesic where the information change is independent of time. 
Note that $\tau(t)$ in Eq. (\ref{eq01}) is related to
the second derivative of the relative entropy  (or Kullback-Leibler divergence) (see Appendix A and \cite{HK16}) and
that ${\cal E}$ is the mean-square  fluctuating energy for a Gaussian PDF (see \cite{KIM16}).

The total change in
information between the initial and final times, $0$ and $t$ respectively, is then 
computed by the total elapsed time in units of $\tau(t)$ as
${\cal{L}} (t)  =  \int_0^{t} \frac{dt_1}{\tau(t_1)}$.
${\cal L}$ (information length) provides the total number of
different states that a system passes through from the initial state with the PDF $p(x,t=0)$ at time $t=0$ 
to the final state with the PDF $p(x,t)$ at time $t$. 
For instance, in equilibrium, $\tau(t_1)$ is infinite so that
measuring $dt_{1}$ in units of this infinite $\tau(t_1)$ at any $t_{1}$ gives $dt_1/\tau(t_1) = 0$
and thus ${\cal L}(t)=0$, manifesting
no flow of time in equilibrium.
See Appendix A for the interpretation of ${\cal L}$
from the perspective of the infinitesimal relative entropy. We note that
${\cal L}$ (and thus $\tau(t)$) is based on Fisher information (c.f.~\cite{fisher}) and is a generalisation of  statistical distance
\cite{WOOTTERS81} to time-dependent problems. 

As an example, let us consider the Gaussian PDF of the total vorticity $\omega_{T} = \langle \omega_{T} \rangle + \omega$ 
given by
\begin{eqnarray}
p(\omega_{T,}\bx,t) 
&=&  \sqrt{\frac{\beta}{\pi}} \exp{[-\beta ( \omega_{T}-\langle \omega_{T} \rangle)^{2}]}.
\label{eq1700}
\end{eqnarray}
Here, the angular brackets denote the average ($\langle \omega \rangle = 0$).
$\beta = \frac{1}{2 \langle \omega^2 \rangle}$ is the inverse temperature and  $\beta \to \infty$ for a very narrow PDF. 
By using the property of the Gaussian distribution (e.g. $\langle \omega^4 \rangle = 3 \langle \omega^2 \rangle ^2$) (e.g. see \cite{POPE,Fokker}), we can show that 
${\cal E}$ in Eq. (\ref{eq01}) is (see \cite{HK16,KH17}).
\begin{eqnarray}
{\cal E}(t)& =& \frac{1}{\tau(t)^{2}}= 
  \frac{1}{2} \frac{(\p_{t}\beta)^{2}}{\beta^{2}} 
+ 2 \beta (\p_{t} \langle \omega_{T} \rangle )^{2 }.
\label{eq003}
\end{eqnarray}
The first term in Eq. (\ref{eq003}) is due to the temporal change in PDF width ($\propto \beta^{{-1/2}}$) while the
second is due to the change in the mean value measured in units of PDF width.

\section{Gaussian PDFs}
To gain a key insight, we start with the case where $\tomega({\bf k}(t=0))$ satisfies the Gaussian statistics. Using Eq. (\ref{eq91}),
we compute the average of the  generating function
$Z = \exp{(i \lambda \omega(x,t))}$ as follows:
\begin{eqnarray}
\langle Z \rangle &= & \langle \exp{(i \lambda \omega(x,t))} \rangle
= \exp{\left [-\frac{1}{2} \lambda^2 \left \langle \omega^{2}(\bx,t)\right \rangle \right]},
\label{eq16}
\end{eqnarray}
where 
\begin{equation}
\left \langle \omega^{2}(\bx,t)\right \rangle =  \int d{\bf k}(t)  d{\bf k'}(t)  e^{i  ({\bf k}(t) + {\bf k}')(t') \cdot {\bf x}} 
\biggl \langle \tomega({\bf k}(t) ) \tomega({\bf k}'(t))\biggr \rangle.
\label{eq17}
\end{equation}
Using Eq. (\ref{eq16}) in Eq. (\ref{eq4}) gives
\begin{eqnarray}
p(\omega,\bx,t) & = &
 \sqrt{\frac{\beta}{\pi}} \exp{[-\beta \omega^{2}]}.
\label{eq170}
\end{eqnarray}
Here, $\beta = \frac{1}{2 \langle \omega^{2}(\bx,t)\rangle}$ is again the inverse temperature. 

On the other hand, taking the time derivative Eq. (\ref{eq16}) and Fourier transform gives
\begin{eqnarray}
\p_t p (\omega, \bx, t) &= &\frac{1}{2}\frac{\partial^2}{\partial \omega^2} \biggl [ [ \p_t  \left \langle \omega^{2}(\bx,t)\right \rangle] p(\omega,\bx,t)\biggr],
\label{eq161}
\end{eqnarray}
consistent with Eq. (\ref{eq170}).\\

In comparing the RHS of Eq. (\ref{eq70}) and Eq.  (\ref{eq161}), we have
\begin{equation}
\frac{1}{2} \p_t  \left \langle \omega^{2}(\bx,t)\right \rangle  = - \nu \langle (\p_j\omega)^2 \rangle  = \nu \langle \omega \nabla^2 \omega \rangle.
  \label{eq162}
 \end{equation}
We will shortly show that Eq. (\ref{eq162}) indeed holds for the Gaussian $\omega(\bx,t)$ in a homogeneous turbulence. In the following subsections, we analyse the zonal case in \S 4.A and the combined shear flow cases $\Omega_z>0$ and $\Omega_s>0$ in \S 4.B, and $\Omega_z>0$ and $\Omega_s<0$ in \S 4.C.

\subsection{ZF case: $\Omega_z>0$ and $\Omega_s=0$}
For the case of zonal flow only (ZF), $\U =  (0, -x\Omega_{z})$, the mean vorticity $\langle \omega_{T}\rangle = -\Omega_{z}$, and the time dependent wavenumber follows from Eq. (\ref{eq9}) as
\begin{eqnarray}
k_x(t) &=& k_x(0) + k_y \Omega_z t,\,\,\,\,\, k_y(t)=k_y(0),
\label{eq10}\\
Q_{1}(t) &\equiv & \int_0^t dt_1 |\bk(t_1)|^2 
=\frac{1}{3} (k_{y} \Omega_{z})^{2} t^{3}+ k_y k_{0 }\Omega_z t^{2}  + (k_x(0)^{2} + k_{y}^{2})t.
\label{eq100}
\end{eqnarray}
With Eq. (\ref{eq100}), Eq. (\ref{eq91}) is rewritten as
\begin{eqnarray}
\tomega(k_x(t),k_y) &=& e^{-\nu Q_{1}(t)} \tomega(k_x(0), k_y).
\label{eq19}
\end{eqnarray}
To compute the mean square of $\omega(\bx,t)$ from Eq. (\ref{eq19}), we assume a homogeneous turbulence at $t=0$ so that the translational invariance in space constrains the correlation function in the Fourier space as
$\biggl \langle \tomega({\bf k}(0) ) \tomega({\bf k}'(0))\biggr \rangle 
=  \delta ({\bf k}(0) + {\bf k}'(0)) \psi ({\bf k}(0))$,
where $\psi({\bf k}(0))$ is the initial power spectrum. Eq. (\ref{eq91}) then give us 
\begin{eqnarray}
\biggl \langle \tomega({\bf k}(t)) \tomega({\bf k}'(t))\biggr \rangle 
&=&  \delta ({\bf k}(0) + {\bf k}'(0)) \psi ({\bf k}(t)),
\label{eq020}\\
\psi({\bk}(t)) &=& e^{ -2 \nu Q_{1}(t) } \psi({\bk}(0)).
\label{eq2000}
\end{eqnarray}
Using Eqs. (\ref{eq020}) and (\ref{eq2000}) in (\ref{eq17}), we then obtain
\begin{eqnarray}
\left \langle \omega^{2}(\bx,t)\right \rangle &=&
\int d{\bk} \psi(\bk(t)).
\label{eq2010}
\end{eqnarray}
We now confirm that Eq. (\ref{eq162}) holds for Eqs. (\ref{eq2000}) and (\ref{eq2010}) since
\begin{equation}
\partial_t \langle \omega^2 \rangle  = -2 \nu \int d\bk |\bk(t)|^2  \psi(\bk(t)),\,\,\
\langle \omega \nabla^2 \omega \rangle = - \int d\bk |\bk(t)|^2  \psi(\bk(t)).
\label{eq202}
\end{equation}
In the following subsections, we discuss PDFs and characteristic dissipation 
times scales by using different $ \psi({\bk}(0))$.
\subsubsection{$\delta$-function power spectrum}
\begin{figure}[h]
\centering
\includegraphics[scale=0.35]{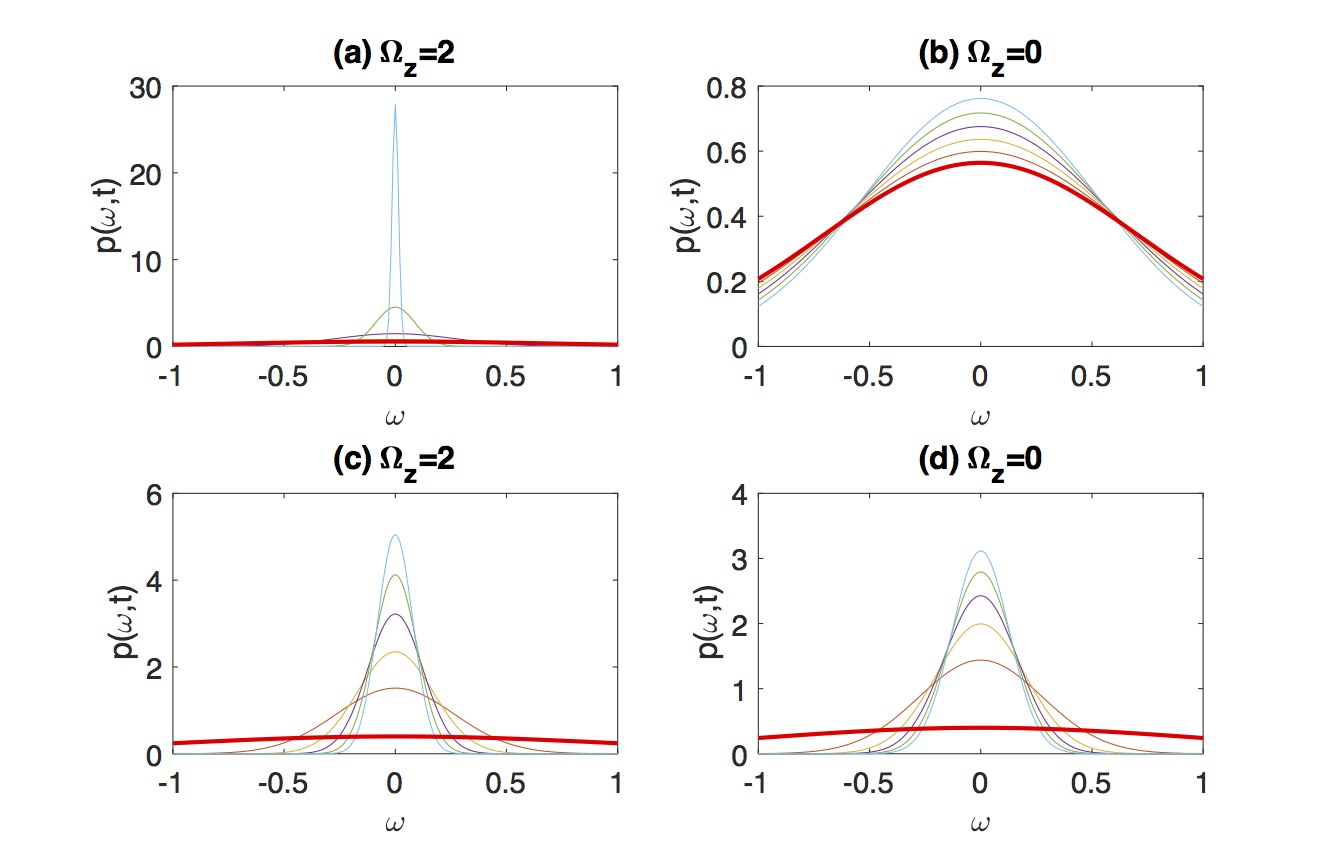}
\caption{Time evolution of $p(\omega,t)$ in panel (a)-(b) for the $\delta$-function power spectrum in (\ref{eq21}) and in panel (c)-(d) for
the Gaussian power spectrum in Eq. (\ref{b1}); $t=0.6\times n$ where $n=0,1,2,3,\cdot\cdot\cdot, 10$ increases from the bottom
to the top curves. The bottom red curve is for the initial PDF. $\alpha=100$, $\nu=0.1$, $k_x(0)=0$, $k_y=1$ and $\phi=1$. }
\vskip 1cm
\label{Fig1}
\end{figure}
The simplest case to consider is a $\delta$-function power spectrum given by:
\begin{equation}
\psi ({\bf k}(0)) = \delta (k_x(0)-a) \delta(k_y-b) \phi,
\label{eq21}
\end{equation}
where $\phi$ is constant. The power spectrum $\psi({\bf k}(t)))$ continues to have a $\delta$-function with the peak at $k_x(t)=a+b\Omega_{z}t$ and $k_y(t)=b$ given by Eq. (\ref{eq10}) with $k_x(0)=a$ and $k_y(0)=b$:
\begin{eqnarray}
\psi({\bk}(t))& = &e^{ -2 \nu Q_{1}(t) } \delta(k_{x}(t)-a-b\Omega_{z}t) \delta(k_{y}(t)-b)\phi.
\label{eq201}
\end{eqnarray}
Using (\ref{eq2010}) and (\ref{eq201}), we have
\begin{eqnarray}
\left \langle \omega^{2}(\bx,t)\right \rangle &=&
 \exp{\left[-\frac{2\nu}{3} \left ((k_{y} \Omega_{z})^{2} t^{3}+ 3k_y \Omega_{z} t^{2} k_{x}(0)  \right) - 2 \nu (k_x({0})^{2} + k_{y}^{2})t \right]} \phi,
\label{eq22}
\end{eqnarray}
providing $\beta = \frac{1}{2 \langle \omega^{2} (\bx,t)\rangle}$ in Eq. (\ref{eq170}).

For a strong shear $\Omega_{z} \gg \nu k_{y}^{2}$, the term $\frac{2\nu}{3} (k_{y} \Omega_{z})^{2} t^{3}$ in Eq. (\ref{eq22}) causes the enhancement of dissipation over a usual exponential viscous damping $\exp{(-2\nu (k_{x}(0)^{2} + k_{y}^{2})t)}$. The effective dissipation time scale $\tau_e$ for such enhanced damping is found from $\frac{2\nu}{3} (k_{y} \Omega_z)^{2} \tau_e^{3} \sim 1$ as
\begin{equation}
\tau_{e} \sim \left(\nu k_{y}^{2}{\Omega_{z}^{2}}\right)^{-{\frac{1}{3}}}
\equiv  \left(\tau_{\eta} \tau_{\Omega_{z}}^{2}\right)^{{\frac{1}{3}}},
\label{eq23}
\end{equation}
where $\tau_{\eta} =\frac{1}{\nu k_{y}^{2}}$ and $\tau_{\Omega_{z}}=\frac{1}{\Omega_{z}}$ are the viscous and shearing time scales, respectively. We now compare $\tau_e$ with the characteristic time scale $\tau(t)= {\cal E}^{{-1/2}}$ in Eq. (\ref{eq003}) over which the information changes. From ${\cal E}$ in Eq. (\ref{eq003}), we have
\begin{equation}
\sqrt{2 {\cal E}(t)}
=\nu |{\bf k}(t)|^{2} 
=  \nu \left(k_{y}^{2} \Omega_{z}^{2} t^{2} + k_{y}k_{x}(0) \Omega_{z}t^{} + k_{x}(0)^{2} + k_{y}^{2} \right).
\label{eq230}
\end{equation}
The time scale $\tau(t)= {\cal E}^{{-1/2}} \propto t^{{-1}}$ as $t \to \infty$ represents a very short dissipation time scale and enhanced dissipation due to the accelerated formation of small scales and their disruption.
Clearly, unlike $\tau_e$, $\tau(t)$ captures the dynamics of the systems, i.e., the dependence of the rate of dissipation on time. When $\Omega_{z}=0$, $\tau(t)=\nu (k_{x}(0)^{2}+k_{y}^{2})$ in Eq.~(\ref{eq230}) becomes constant, which is the case of a geodesic (see \S 3). The value of this constant $\tau(t)$ however depends on initial wave number ${\bk}(0)$, meaning that $\tau(t)$ is not scale invariant. This is to be contrasted to the case considered in \S 4.B.2. Scalings of $\tau_e$ and $\tau(t)$ are summarized in Table~I. 
\begin{table}[b]
\begin{center}
\begin{tabular}{ |l || c || c |  }
\hline
{Shear flows} & ZF: $\Omega_{z}$ &ZF+ST: $\Omega_{z}=\Omega_{s}=\Omega$ \\
\hline \hline
         \,\,$\delta$-function  & $\tau_{e} \propto \Omega_{z}^{-\frac{2}{3}}$  
                & $\tau_{e} \propto \Omega^{-1} \ln{\Omega}$     \\
              \,\,spectrum                  &  $ \tau(t) \propto t^{{-1}}$  
                & $\tau(t) \propto  e^{-\Omega t}$      \\
\hline
             \,\,Constant  & $\tau_{e} \propto \Omega_{z}^{-\frac{1}{2}}$ & $\tau_e \propto \Omega^{-1} $    \\
                \,\,spectrum   &$ \tau(t) \propto t $  & $ \tau(t) \propto [\Omega\, {\tanh{(\Omega t)}}]^{-1} $    \\
\hline
\end{tabular}
\caption{\label{table1}Scalings of $\tau_e$ and $\tau(t)$ for the initial $\delta$-function and constant power spectra
in the case of ZF with shearing rate $\Omega_z$ and hyperbolic ZF+ST with shearing rate $\Omega_z=\Omega_s=\Omega$.
Gaussian power spectrum has the scaling between $\delta$-function and constant power spectra.}
\end{center}
\end{table}

Figure 1(a)-(b) compares the time evolution of $p(\omega,t)$ for $\Omega_{z}=2$ in (a) and for $\Omega_z=0$ in (b) by using $k_{y}=1$, $k_x(0)=0$, $\phi=1$ and $\nu=0.1$.
The initial PDF is shown in the bottom red curve and the time increases from the bottom to the top curve as $t=0.6\times n$ where
 $n=0,1,2,3,\cdot\cdot\cdot, 10$. The narrowing of PDF width in time in Figure 1(a) is in sharp contrast to a much smaller change in Figure 1(b) between $t=0$ and $t=6$.
A much faster narrowing in Figure 1(a) manifests the enhanced dissipation of the mean square vorticity by $\Omega_z$. 
\subsubsection{Constant power spectrum}
$\tau_e\propto \Omega_z^{-2/3}$ in Eq. (\ref{eq23}) is specific to the case of the $\delta$-function power spectrum where there is unique wavenumber at $t=0$ that evolves according to Eq. (\ref{eq10}). To understand how $\tau_{e}$ is affected in the presence of different $\bk(0)$ modes,
we consider a constant spectrum by taking  
$\psi = {\rm constant} = \phi$. Then, the power spectrum evolves in time as follows:
\begin{eqnarray}
\psi({\bk}(t),t)& = &  e^{ -2\nu Q_{1}(t) } \phi,
\label{eq01900}
\end{eqnarray}
where $Q_{1}(t)$ is given in Eq. (\ref{eq100}).
From Eqs. (\ref{eq2010}) and (\ref{eq01900}), we obtain
\begin{eqnarray}
\left \langle \omega^{2}(x,t)\right \rangle &=&
\frac{\pi}{ 2 \nu t \sqrt{4 + \frac{1}{3} \Omega_{z}^{2}t^{2}}} \phi.
\label{eq24}
\end{eqnarray}
after performing the Gaussian integrals over $k_{y}$ and $k_x(0)$. Eq. (\ref{eq24}) shows that $\Omega_z$ changes the scaling of 
$\left \langle \omega^{2}(x,t)\right \rangle$ from $t^{-1}$ to $t^{{-2}}$ by the enhanced dissipation. In this case, $\tau_e$ is found from $\nu \Omega_{z} \tau_e^{2} \sim 1$ as
\begin{equation}
\tau_{e} \propto (\Omega_{z} \nu)^{{-\frac{1}{2}}}.
\label{eq25}
\end{equation}
Thus, the dependence of $\tau_{e} \propto \Omega_{z}^{{-1/2}}$ on $\Omega_z$ is weaker than $\tau_{e} \propto \Omega_{z}^{{-2/3}}$ for a $\delta$-function spectrum as the effect of shearing is reduced
in the case of multiple modes. This is basically because the distortion of an eddy by shearing follows a wave number specific time evolution (e.g. Eq. (\ref{eq10})); the effect of a shear on multiple modes is not coherent as eddies with different wave numbers evolve differently, and is thus less 
effective.  

This reduced shearing effect can also be inferred from
${\cal E}$ in Eq. (\ref{eq003}), which becomes
\begin{equation}
{\cal E}(t)= \frac{(4 + \frac{2}{3} \Omega_{z}t^{2} )^{2} }{2 t^{2}( 4 +\frac{1}{3} \Omega_{z}t^{2} )^{2} } .
\label{eq250}
\end{equation}
Thus, Eq.  (\ref{eq250}) gives the time scale $\tau(t)= {\cal E}^{{-1/2}} \propto t$ for $\Omega_{z}t \gg 1$. 
This should be compared
with $\tau(t) \propto t^{-1}$ in the case of a $\delta$-function power spectrum above (see Table I). The increase of $\tau(t)$ with $t$ means a longer time scale of dissipation and thus it
manifests that the dissipation becomes less effective for large time.
Similar results are shown for the case of an anisotropic power spectrum with $k_x(0) = 0$ in Appendix B. 
We note that the mean square vorticity in Eq. (\ref{eq24}) apparently diverges at $t=0$ due to the unlimited range of the $\bk$ integral for
an initial constant power spectrum. Mathematically, this problem is readily ratified by using a localised spectrum in the next subsection.


\subsubsection{Gaussian power spectrum}
The initial Gaussian power spectrum
$\psi(0) = \frac{1}{\alpha \pi} e^{-\frac{1}{\alpha} (k_x(0)^{2} + k_{y}^{2}) } \phi_{}$ gives
\begin{eqnarray}
\psi({\bk}(t))& = &\frac{1}{\alpha \pi} e^{-\frac{1}{\alpha} (k_x(0)^{2} + k_{y}^{2}) -2 \nu Q_{1} (t)}\phi_{},
\label{eq0190}
\end{eqnarray}
where $Q_{1}(t)$ is given in Eq. (\ref{eq100});
$\alpha$ represents the width of the initial power spectrum; $\alpha \to 0$ and $\alpha \to \infty$ recovers the $\delta$-function and constant
power spectrum, respectively. To understand the effect of $\Omega_z$ on the evolution of $\psi({\bk}(t))$ in (\ref{eq0190}), we use $k_x(0) = k_x(t) - \Omega_z k_y t$ and present $\psi({\bk}(t))$ in Figure 2. Without diffusion ($\nu=0$), $\psi(\bk(t))$ in Figure 2(a) shows the generation of large $k_x$ wave number due to ZF shearing. When a diffusion ($\nu=0.1$) is included in Figure 2(b), large $k_x$ (and $k_y$) modes quickly damp due to molecular dissipation, $\psi({\bk}(t))$ forming a sharp peak around $k_x=k_y=0$.
\begin{figure}
    \centering
    \subfigure[]
    {
        \includegraphics[scale=.22]{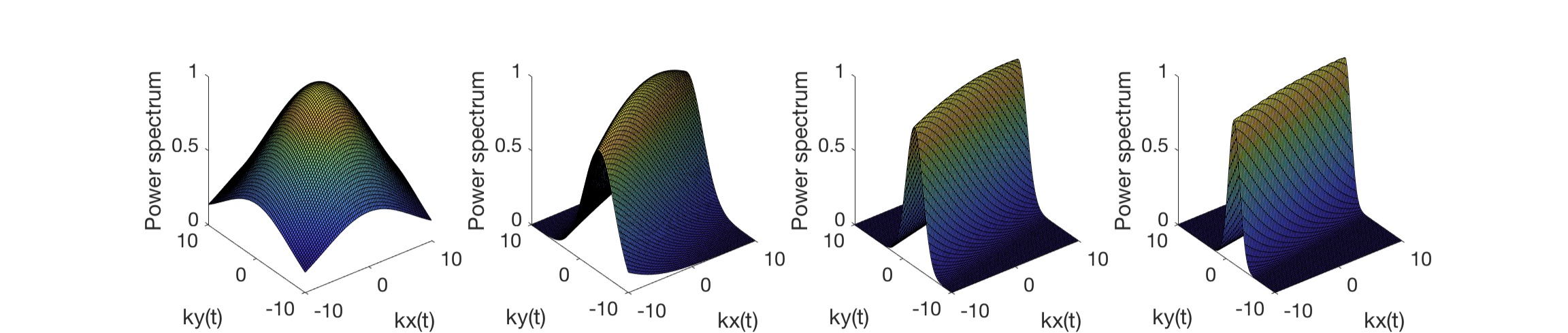}
    }
    \\
    \subfigure[]
    {
        \includegraphics[scale=.22]{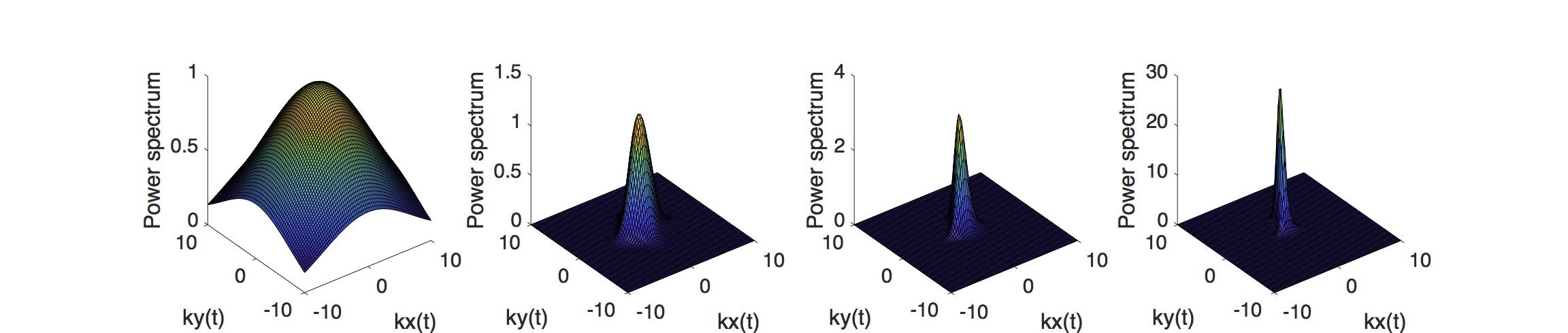}
    }
        \caption
    {
        (a) Time evolution of power spectrum ($t=0,1,2,3$ increasing from left to right) for  $\Omega_z= 2$, $\Omega_s=0$, $\alpha=100$, $\phi=1$, $\nu=0.$ (b) The same as (a)  but for $\nu=0.1$.
    }
    \label{fig:Fig2}
\end{figure}

On the other hand, using Eqs. (\ref{eq0190}) and (\ref{eq100}) in (\ref{eq2010}), we find  
\begin{eqnarray}
\left \langle \omega^{2}(x,t)\right \rangle
 &=&  \frac{1}{\alpha} \sqrt{\frac{1}{A}} \phi_{},
\label{bb1}
\end{eqnarray}
where $A =  (2 \nu t + \alpha^{-1})^{2} + \frac{1}{3} (\nu t) (\Omega_{z} t)^{2} [\nu t + 2 \alpha^{-1}]$.
We note that in the limit of $t \to 0$ and $t \to \infty$, Eq. (\ref{b1}) is reduced to
\begin{eqnarray}
 \left \langle \omega^{2}(x,t)\right \rangle_{t \to 0} \to \phi_{}, \,\,&&
\left \langle \omega^{2}(x,t )\right \rangle_{t \to \infty} \to\frac{\sqrt{3} }{\alpha \nu \Omega_{z} t^{2}} \phi_{}, 
\label{bb3}
\end{eqnarray}
respectively. The second equation in Eq. (\ref{bb3}) recovers the limit of a $\delta$-function power spectrum in Eq. (\ref{eq24})
(up to an unimportant small numerical factor).
Eq. (\ref{bb1}) very conveniently shows the transition of the scaling of $\tau_e$ from $\propto \Omega_z^{-2/3}$ in Eq. (\ref{eq23}) to $\propto \Omega_z^{-1/2}$ in Eq. (\ref{eq25}) as $\alpha$ is increases (see also above). 

Figure 1(c) and 1(d) show the evolution of $p(\omega,t)$ for this Gaussian power
spectrum for $\Omega_z=2$ and $\Omega_z=0$, respectively. Here, parameter values are the same as those in Figure 1(a)-(b) apart from
$\alpha = 100$.
Comparing Figure 1(c) with Figure 1(a), we see much slower narrowing of the PDFs as the shearing effect is less effective in the presence
of multiple ${\bk}$ modes. As observed in Figure 1(a)-(b), the PDF in Figure 1(d) for $\Omega_z=0$ narrows slower than that in Figure 1(c). However, comparing Figure 1(b) and 1(d), the presence of multiple ${\bk}$ modes tends to promote dissipation (due to high wave number modes).

\subsection{Hyperbolic ZF+ST case: $\Omega_{z}>0$ and $\Omega_{s}>0$}
Compared with the case of zonal flows, the combined effect of Zonal Flows and STreamers (ZF+ST) have been studied much less. We show below that the action of ZF+ST can lead to an exponentially fast formation of small scale structure. For  
${\U} = (-y \Omega_s, -x \Omega_z)$ with $\Omega_s>0$ and $\Omega_z>0$, $\U$ has the mean vorticity $\nabla \times \U = (-\Omega_z + \Omega_s) {\hat z}$, which becomes zero for $\Omega_z = \Omega_s$.
The solution to Eq. (\ref{eq10}) can be found as:
\begin{eqnarray}
k_y(t) =
{\ok}  \frac{\Omega}{\Omega_z}\cosh{(\Omega t + \theta)},\,\,
&&  k_x(t) =
 {\ok} \sinh{(\Omega t + \theta)},
\label{eq12}
\end{eqnarray}
where
\begin{eqnarray}
&& \Omega =  \sqrt{\Omega_z \Omega_s},\,\,\,
{\ok}  \left[ k_y(0)^2 + k_x(0)^2 \frac{\Omega^2}{\Omega_z^2} \right]^{\frac{1}{2}},
\label{eq14} \\
&& \sinh{(\theta)} = \frac{k_x(0)}{\ok},\,\,\,\, \cosh{\theta} = \frac{\Omega_z k_y(0)}{\Omega \ok}.
\label{eq15}
\end{eqnarray}
We focus on the case of $\Omega_{z}=\Omega_{s}=\Omega$ with zero mean vorticity, in which case $k_{x}^{2}+k_{y}^{2} =
{\ok}^{2} \cosh{[2 (\Omega t + \theta)]}$ follows from Eq. (\ref{eq12}). Thus, with the help of Eqs. (\ref{eq14})-(\ref{eq15}), we obtain
$Q_{2} (t) = \int_0^t dt_1 |\bk(t_1)|^2$ as 
\begin{eqnarray}
Q_{2} (t) 
&=& \frac{1}{4 \Omega} \left[ (k_{x}(0) + k_{y}(0))^{2} (e^{2 \Omega t}-1) + (k_{x}(0) - k_{y}(0))^{2} (1-e^{-2 \Omega t}) \right].
\label{eq26}
\end{eqnarray}
Since ${\bf k}(t)$ starting with ${\bk}(0)$ changes in time according to Eq. (\ref{eq12}), in order to see how the power spectrum evolves
in time, we need to express $Q_2(t)$ in Eq. (\ref{eq26}) in terms of ${\bk}(t)$. To this end, we solve Eq. (\ref{eq12}) for $k_x(0)$ and $k_y(0)$ to find
$k_y(0) = k_y(t) \cosh{(\Omega t)} - k_x (t) \sinh{(\Omega t)}$ and
$k_x(0) =  k_x(t) \cosh{(\Omega t)} - k_y(t) \sinh{(\Omega t)}$,
and thus
\begin{eqnarray}
k_x(0) + k_y(0) &= &[k_x(t) + k_y(t)] e^{-\Omega t} ,
\nonumber\\
k_x(0) -k_y(0) &=& [k_x(t) - k_y(t)] e^{\Omega t}.
\label{eq122}
\end{eqnarray}
By using Eq. (\ref{eq122}) in Eq. (\ref{eq26}), we have
\begin{eqnarray}
Q_2(t) &=&
\frac{1}{4\Omega} \biggl[ [k_x(t) + k_y(t)]^2  (1- e^{-2 \Omega t})
+ [k_x(t) - k_y(t)]^2  (e^{ 2\Omega t}-1)\biggr].
\label{eq124}
\end{eqnarray}
Interestingly, Eq.  (\ref{eq124}) shows that the dissipation $Q_2(t)$ takes its minimum value when $k_x(t) = k_y(t)$. Furthermore, from Eq. (\ref{eq122}), we also find  
\begin{eqnarray}
 k_x(0)^2 + k_y(0)^2 
&= &\frac{1}{2} \left[ (k_x(t)-k_y(t))^2 e^{2 \Omega t} + (k_x(t)+k_y(t))^2 e^{-2\Omega t}\right],
 \label{eq1240}
\end{eqnarray}
which also takes its minimum along $k_x(t) = k_y(t)$. The minimum of  Eqs. (\ref{eq124}) and 
(\ref{eq1240}) along $k_x(t) = k_y(t)$ is later shown to give a peak in the power spectrum $\psi({\bk} (t))$ in \S 5 (see Figure 3).
We refer to $k_x(t) = k_y(t)$ as the principle direction in the following.


\subsubsection{$\delta$-function power spectrum}
For a $\delta$-function power spectrum given by Eq. (\ref{eq21}), the power spectrum $\psi({\bk}(t))$ continues to have a $\delta$-function with the peak at $k_x(t)$ and $k_y(t)$ given by Eq. (\ref{eq12}) with $k_x(0)=a$ and $k_y(0)=b$. This leads to
\begin{eqnarray}
\left \langle \omega^{2}(\bx,t)\right \rangle &=&
 \exp{\left[-\frac{\nu}{2\Omega }\left[  (k_{x}(0) + k_{y}(0))^{2} (e^{2 \Omega t}-1) - (k_{x}(0) - k_{y}(0))^{2} (e^{-2 \Omega t}-1) \right] \right]} \phi.
\label{eq27}
\end{eqnarray}
From Eq. (\ref{eq27}), we find the effective diffusion time $\tau_e$
\begin{equation}
\tau_e \sim \frac{1}{2 \Omega}  \ln{\left( \frac{2 \Omega}{\nu |{\bk}_{0}|^{2}} \right) }.
\label{eq28}
\end{equation}
$\tau_e$ in Eq. (\ref{eq28}) is smaller than Eq. (\ref{eq23}) for a sufficiently large $\Omega$, with a stronger dependence $\Omega ^{{-1}}\ln{\Omega}$
on $\Omega$,
in comparison with $\Omega_{z}^{-2/3}$ in ZF case. 
\begin{figure}
\centering
\includegraphics[scale=0.35]{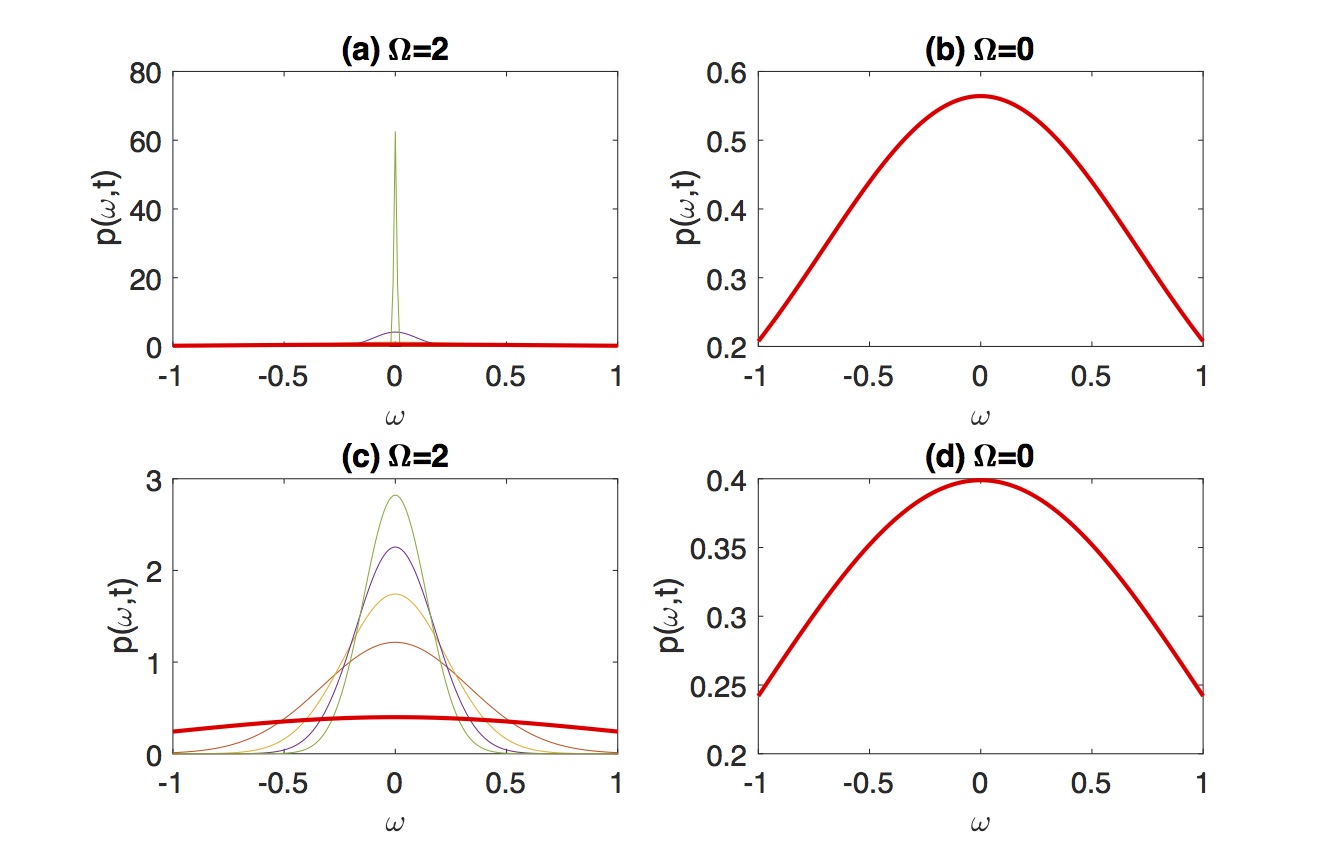}
\caption{Time evolution of $p(\omega,t)$ for the $\delta$-function power spectrum in Eq. (\ref{eq27}) in panel (a)-(b) and for
the Gaussian power spectrum in Eq. (\ref{eq191}) in panel (b)-(d); $t=0.2\times n$ where $n=0,1,2,3,4,5$ increases from the bottom
to the top curves. The bottom red curve is for the initial PDF. $\alpha=100$, $\nu=0.1$ and $\phi=1$.}
\vskip 1cm
\label{Fig3}
\end{figure}
Furthermore, due to the double exponential decrease in $\langle \omega^2 \rangle$, $\tau(t)$ in Eq. (\ref{eq003}) is reduced exponentially fast as: 
\begin{equation}
\tau(t)\propto e^{{-\Omega t}}.
\label{eq2300}
\end{equation}
The exponentially decreasing $\tau(t)$ in Eq. (\ref{eq2300}) reflects a very efficient dissipation by ZF+ST.
The evolution of PDF is shown in Figure 3 for $\Omega=2$ in (a) and  $\Omega=0$ in (b).
The time $t=0.2\times n$ where $n=0,1,2,3,4,5$ increases from the bottom
to the top curves. The bottom red curve is for the initial PDF. Comparing Figure 3(a) with
Figure 1(a), we see a much faster narrowing of the PDFs in the hyperbolic ZF+ST case due to a much faster dissipation.
(Note that the total time span $t=[0,1]$ in Figure 3 is much smaller than $t=[0,6]$ in Figure 1.) The change in Figure 3(b) with $\Omega=0$
is too small to be seen.

\subsubsection{Constant power spectrum}
For an initial constant power spectrum $\psi(0)=\phi$, the power spectrum again evolves as
$\psi({\bk}(t),t) = e^{ -2 \nu Q_{2}(t) } \phi$,
where $Q_2$ is given in  Eq. (\ref{eq26}).
Therefore, by using Eqs. (\ref{eq2010}) and (\ref{eq26}), we find  
\begin{eqnarray}
&&\left \langle \omega^{2}(x,t)\right \rangle
 = \frac{1}{2} \int dp dq  
\exp{\left[-\frac{\nu}{2\Omega } \left[ p^{2} (e^{2 \Omega t}-1) + q^{2} (1-e^{-2 \Omega t}) \right] \right]} \phi
 = \frac{ \Omega\phi}{2 \pi \nu \sinh{(\Omega t)}}.
  \label{eq29}
\end{eqnarray}
Here, we performed the integrals over $p\equiv k_{x}(0)+k_{y}(0)$ and 
$q\equiv k_{x}(0)-k_{y}(0)$.

Compared with the $\delta$-function power spectrum, the effect of shear flow is reduced from double exponential to exponential. 
For $\Omega t \gg 1$, $\langle \omega^{2}(x,t) \rangle\sim \frac{ \Omega}{ 2 \pi \nu} e^{-\Omega t}\phi$, giving
an effective diffusion time 
\begin{equation}
\tau_e \sim {\Omega}^{-1}.
\label{eq290}
\end{equation}
Interestingly, $\tau(t)$ in this case has a similar dependence on $\Omega$ since
\begin{equation}
\tau(t) =\frac{1}{ \Omega \tanh{(\Omega t)}},
\label{eq2900}
\end{equation}
approaching a constant value $ \Omega^{-1}$ (!) for $t \gg {\Omega}^{-1}$. This is another example of a geodesic, which is
more interesting than the case of $\Omega_{z}=0$ in Eq. (\ref{eq230}) because Eq. (\ref{eq2900})
is induced by non-zero $\Omega$ in the presence of different ${\bk}(t)$ modes which evolve from an initial constant power spectrum. 
In fact, $\tau(t) \sim \Omega^{-1}$ explicitly shows that $\Omega$ is the very cause of information change.
On the other hand, in comparison with the exponentially decreasing $\tau(t)$ in Eq. (\ref{eq2300}), Eq. (\ref{eq2900}) again
illustrates the reduced shearing effect due to the presence of multiple ${\bk}$ modes.
Finally, we note that the divergence at $t=0$ is due to the unbounded power spectrum
as in the case of Eq. (\ref{eq24}). Scalings of $\tau_e$ and $\tau(t)$ are summarized in Table I. 

\subsubsection{Gaussian power spectrum}

For
$\psi(0) = \frac{1}{\alpha \pi} e^{-\frac{1}{\alpha} (k_x(0)^{2} + k_{y}(0)^{2}) } \phi_{}$,
we have
\begin{eqnarray}
\psi(\bk(t))& = &\frac{1}{\alpha \pi} e^{-\frac{1}{\alpha} (k_x(0)^{2} + k_{y}(0)^{2}) -2 \nu Q_{2}(t)}\phi_{},
\label{eq190}
\end{eqnarray}
where $Q_{2}$ is given by Eq. (\ref{eq26}). By using Eqs. (\ref{eq124}) and (\ref{eq1240}) in Eq. (\ref{eq190}), we present
the evolution of the power spectrum $\psi(\bk(t))$ in Figure 4 for $\Omega=2$, where time increases from left to right as $t=0, 0.6,1.2,1.8$. 
Without diffusion ($\nu=0$), $\psi(\bk(t))$ in Figure 4(a) shows a fast reduction in $\psi({\bk}(t))$ along $k_x(t)+k_y(t)=0$, with the peak
forming along the principle direction $k_x(t)=k_y(t)$. When diffusion ($\nu=0.1$) is included in Figure 4(b), modes of large wavenumber
also damp along the principle direction in time due to the molecular dissipation although the damping is weaker compared to
that along $k_x(t)+k_y(t)=0$.
This is because the dissipation $Q_2(t)$
in Eq. (\ref{eq124}) and $k_x(0)^2 + k_y(0)^2$ in Eq. (\ref{eq1240}) are minimized along $k_x(t)=k_y(t)$,
as noted previously.

\begin{figure}
    \centering
    \subfigure[]
    {
        \includegraphics[scale=.22]{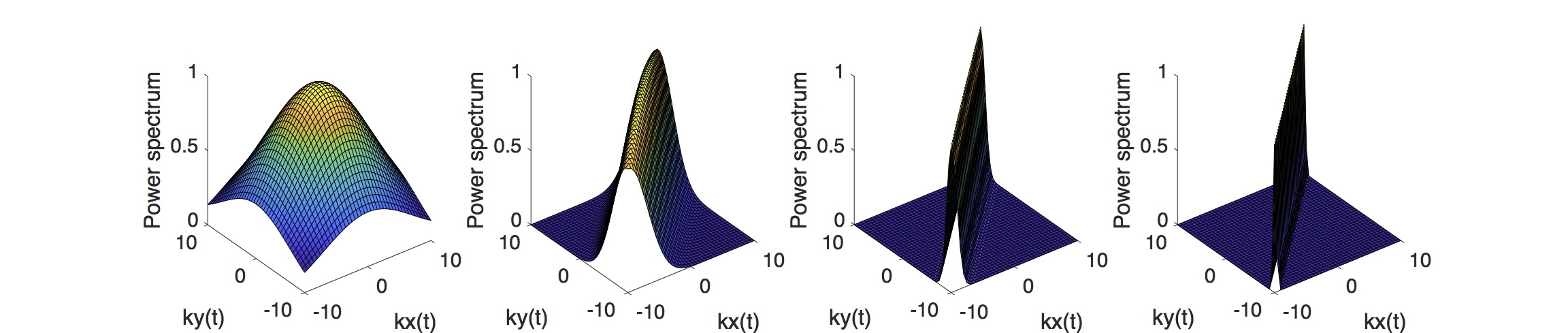}
    }
    \\
    \subfigure[]
    {
        \includegraphics[scale=.22]{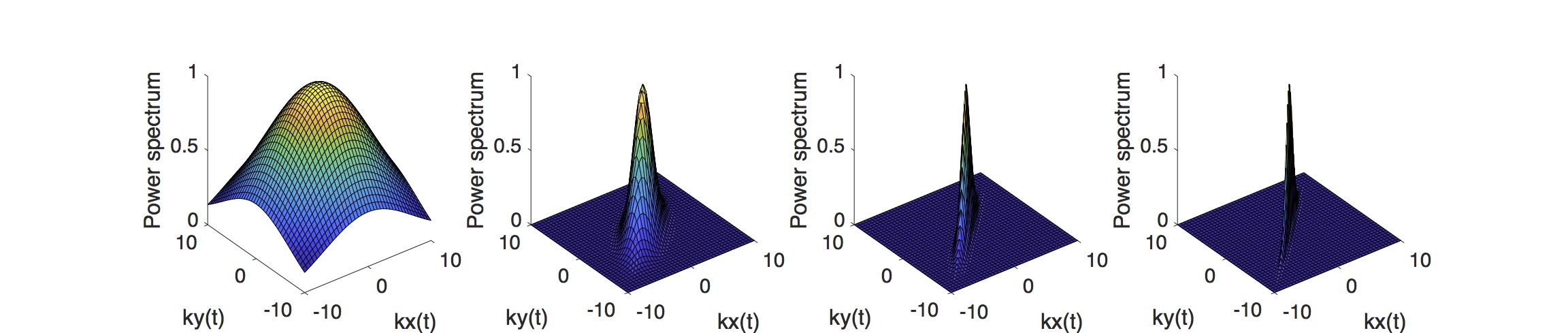}
    }
    \qquad
        \caption
    {
        (a) Time evolution of power spectrum ($t=0, 0.6, 1.2,1.8$ increasing from left to right) for  $\Omega_z= \Omega_s=\Omega= 2$, $\alpha=100$, $\phi=1$, $\nu=0.$ (b) The same as (a) but for $\nu=0.1$.
    }
    \label{Fig4}
\end{figure}



Now, Eq. (\ref{eq190}) leads to the mean square vorticity:
\begin{eqnarray}
\left \langle \omega^{2}(x,t)\right \rangle
&=& \frac{2}{\alpha \pi} 
\int dp dq  
\exp{\left[-\frac{\nu}{2\Omega } \left[ p^{2} (e^{2 \Omega t}-1) + q^{2} (1-e^{-2 \Omega t})-\frac{1}{\alpha} (p^2+q^2) \right] \right]}
 \phi
 \nonumber \\
 &&= \frac{2}{\alpha \sqrt{A B}},
  \label{eq191}
\end{eqnarray}
where
\begin{eqnarray}
A = \frac{\nu}{2 \Omega} (e^{2 \Omega t} -1) + \frac{1}{\alpha},\,\,
&&
B = \frac{\nu}{2 \Omega} (1-e^{-2 \Omega t} ) + \frac{1}{\alpha}.
\label{192}
\end{eqnarray}
$\tau_{e} \propto {\Omega}^{-1}$ is thus similar to Eq. (\ref{eq290}) for $t \gg \Omega^{-1}$. However, in contrast
to Eq. (\ref{eq2900}), $\tau(t)$ becomes constant for $t \gg \Omega^{-1}$ only for
a sufficiently large $\alpha$, that is, in the limit of a constant power spectrum.
Figure 3(c) shows the evolution of $p(\omega,t)$ for this case using the same parameter values $\Omega_z=\Omega_s=\Omega=2$ as in Figure 3(a) apart from
$\alpha = 100$.
Comparing Figure 3(c) with Figure 3(a), we see much slower narrowing of the PDFs as the shearing effect is less effective in the presence
of multiple ${\bk}$ modes, as observed in Figure 1. The evolution of $p(\omega,t)$ for $\Omega=0$ is shown in Figure 3(d), which hardly changes.

\subsection{Elliptic ZF+ST case}
For the hyperbolic ZF+ST case in \S 4.B, the sign of zonal flow and streamer shear is the same. When they have different sign, ZF+ST leads to a
rotating wave number. To see this, we consider
${\U} = (y \Omega_s, -x \Omega_z)$ with $\Omega_s>0$ and $\Omega_z>0$ which has the non zero mean vorticity $\nabla \times \U =- (\Omega_z + \Omega_s) {\hat z}$. For this ZF+ST, we find the solution to Eq. (\ref{eq10}) as 
$k_y(t)  = {\ok}\frac{\Omega}{\Omega_z}   \cos{(\Omega t + \theta)}$ and
$k_x(t) =
{\ok} \sin{(\Omega t + \theta)}$,
where
$\Omega =  \sqrt{\Omega_z \Omega_s}$,
${\ok} = \left[ k_x(0)^2 + k_y(0)^2 \frac{\Omega_{z}^2}{\Omega^2} \right]^{\frac{1}{2}}$, and
$\sin{(\theta)} =\frac{k_x(0)}{\ok},\,\,\,\, \cos{\theta} = \frac{\Omega_z k_y(0)}{\Omega \ok}$.
When $\Omega_{z}=\Omega_{s}=\Omega$, $k_{x}^{2}+k_{y}^{2} =k_{x}(0)^{2}+ k_{y}(0)^{2}$ is constant in time, with no enhancement of
dissipation. However, for $\Omega_{z} \ne \Omega_{s}$, 
$k_{x}^{2}+k_{y}^{2 }= {\ok}^{2} \left[ \sin^{2}{(\Omega t + \theta)} + \frac{\Omega_{s}^{}}{\Omega_{z}^{}} \cos^{2}{(\Omega t + \theta)}\right]$.
Although the overall dissipation may not be significantly enhanced by this shear flow, there is an interesting effect on the dynamics due
to oscillatory dissipation, $k_x$ or $k_y$, which provides a periodic background (or potential). This is discussed in our accompanying paper \cite{MK17}.

\section{Non-Gaussian PDFs}
In the previous section, we investigated the effect of shear flows on the evolution of the Gaussian PDFs and power spectra. The main effect was the shift of power to larger wavenumber, accelerating dissipation and narrowing PDF width. We now extend our study to a non-Gaussian case to examine the effect of shear flows on the form of PDF. Although there are many possible causes for non-Gaussian PDFs, we consider one example of an inhomogeneous turbulence. That is, we drop the assumption of homogenous turbulence and instead prescribe the profile of the initial vorticity fluctuation as
\begin{equation}
\tomega(\bk(0),t=0) = \frac{1}{\alpha \pi} e^{-\frac{1}{\alpha} (k_x(0)^{2} + k_{y}(0)^{2}) },\,\,\,\,
\omega(\bx,t=0) = e^{-\frac{\alpha}{4} (x^{2} + {y}^{2})},
\label{eq40}
\end{equation}
where $\alpha$ is a positive random variable. Note that when $\alpha=0$, Eq. (\ref{eq40}) gives a constant $\omega(\bx,0)$ while non zero constant
$\alpha$ ($>0$) gives the typical length scale $l$ of the profile of the initial vorticity fluctuation as $l \sim \alpha^{-1/2}$. A random
positive $\alpha$ makes the profile of the initial vorticity fluctuation on different length scales. 
By considering the hyperbolic shear flow considered in \S 4.B, we have
\begin{eqnarray}
\tomega(\bk(t),t)& = &\frac{1}{\alpha \pi} e^{-\frac{1}{\alpha } (k_x(0)^{2} + k_{y}(0)^{2}) -\nu Q_{2}(t)},
\label{eq41}
\end{eqnarray}
where $Q_{2}(t)$ is given in Eq. (\ref{eq26}). In order to take the inverse Fourier transform of Eq. (\ref{eq41}) to find 
$\omega(\bx,t)$, we first write Eq.  (\ref{eq12}) in terms of  $p=k_{x}(0) + k_{y}(0)$ and
$q = -k_{x}(0) +  k_{y}(0)$ as
$k_y(t) = \frac{1}{2} \left [ p e^{\Omega t} + q e^{-\Omega t} \right ]$ and
$k_x(t) =\frac{1}{2} \left [ p e^{\Omega t} - q e^{-\Omega t} \right ]$ 
so that 
\begin{equation}
{\bk}(t) \cdot \bx   = \frac{1}{2} \left [ p e^{\Omega t} z_{1}+ q e^{-\Omega t} z_{2}\right ],
\label{eq44}
\end{equation}
where
\begin{equation}
z_{1} = x+y, \,\,\,\, z_{2} = y-x.
\label{eq45}
\end{equation}
Then, by using Eqs. (\ref{eq26}), (\ref{eq41}) and (\ref{eq44}), we obtain $\omega(\bx,t) =\int d{\bk}(t)\, e^{i{\bk}(t) \cdot \bx } \tomega(\bk(t),t)$ as
\begin{eqnarray}
&&\omega(\bx,t) 
= \frac{1}{2 \alpha  \sqrt{C D}} \exp{\left [-\frac{e^{2 \Omega t} z_{1}^{2}}{8 C}-\frac{e^{-2 \Omega t} z_{2}^{2}}{8 D}\right]}.
  \label{eq46}
\end{eqnarray}
Here
\begin{eqnarray}
C= \frac{\nu}{4 \Omega} (e^{2 \Omega t} -1) + \frac{1}{2 \alpha},\,\,
&&
D = \frac{\nu}{4 \Omega} (1-e^{-2 \Omega t} ) + \frac{1}{2 \alpha}.
\label{eq47}
\end{eqnarray}
As $t \to 0$, $C \to \frac{1}{2 \alpha}$, $D \to \frac{1}{2 \alpha}$, and Eq. (\ref{eq46}) recovers Eq. (\ref{eq40}).
For $t \ne 0$, $C$ and $D$ depend on the relative magnitude of $4\nu/\Omega$ and $1/2 \alpha$.
\begin{figure}[t]
\centering
\includegraphics[scale=0.22]{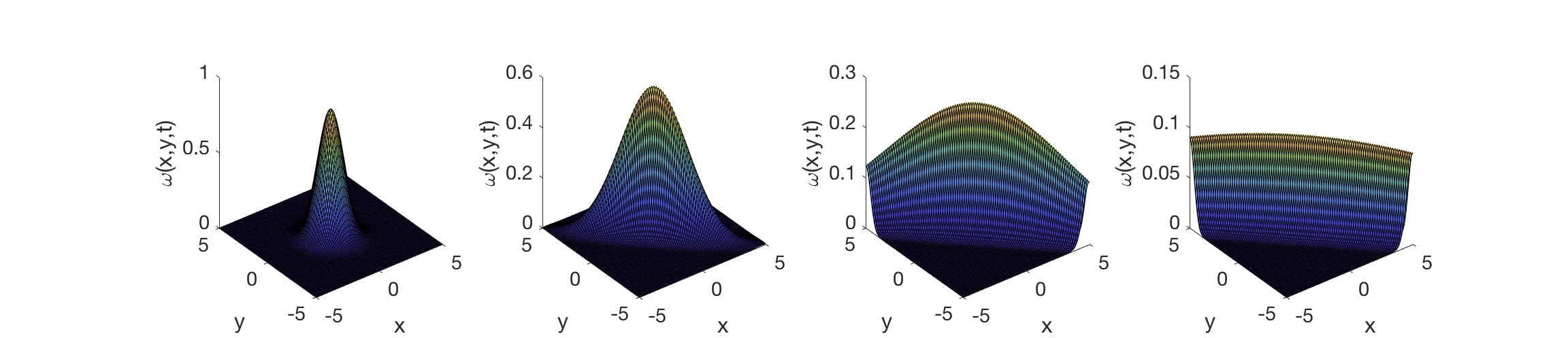}
\caption{Time evolution of $\omega(x,y,t)$ for $\Omega_z= \Omega_s=\Omega=2$, $t=0, 0.5, 1, 1.5$ increasing from left to right;
$\alpha=2$, $\nu=0.1$
and $\phi=1$. }
\vskip 1cm
\label{Fig5}
\end{figure}

Before proceeding to random $\alpha$, we note that for a constant values of $\alpha$, Eq. (\ref{eq46}) shows the anisotropic distortion and decay of the profile of vorticity fluctuation by shear flows. The time evolution of $\omega(\bx,t)$ for constant $\alpha =100$
is shown in Figure 5, where time $t=0, 5, 1, 1.5$ increases from left to right. Of notable is the flattening and elongation of $\omega(x,y,t)$ along $z_1=x+y=0$, with
the formation of a sheet like structure. This is quite similar to what is seen in Figure 4, recalling 
that a narrow ${\bk}$ profile corresponds to a broad $\bx$ profile. 

When $\alpha$ ($>0$) is random, the statistics of $\omega(\bx,t)$ depends on $\alpha$ as
\begin{eqnarray}
p(\omega,\bx, t) &=& \left | \frac{d \alpha}{d \omega} \right | p(\alpha).
\label{eq49}
\end{eqnarray}
In particular, at $t=0$, Eq. (\ref{eq40}) gives $\alpha = -4 \ln{(\omega(t=0))}/r^{2}$ where $r^{2}=x^{2}+y^{2}$, leading to
\begin{eqnarray}
p(\omega,\bx,0) &=&
 \frac{4}{ \omega r^{2}} p(\alpha).
\label{eq50}
\end{eqnarray}

For our purpose, it suffices to assume that $\alpha$ is uniformly distributed within a certain range. Two cases of our interest is the limit of weak inhomogeneity where i) $\alpha= [0, \frac{2 \Omega}{\nu}e^{-2 \Omega t}]$ and of a strong inhomogeneity { where ii) $\alpha=[ \frac{2 \Omega}{\nu}, \alpha_c]$ with $\alpha_c > \frac{2 \Omega}{\nu}$}. 
In case i), the shearing does not have much influence on the scale of inhomogeneity while in case ii), it does have a significant effect.
Starting our analysis in case i), we approximate $C \sim  D \sim \frac{1}{2 \alpha}$, and consequently
\begin{equation}
\omega(\bx,t) \sim \exp{\left[-\frac{\alpha}{4} \left(e^{2 \Omega t} z_1^2 + e^{-2\Omega t} z_2^2\right) \right]}
= \exp{\left[-\frac{\alpha}{4} G_{1}\right]},
\label{eq51}
\end{equation}
where $G_{1} = e^{2 \Omega t} z_1^2 + e^{-2\Omega t} z_2^2$. Eqs. (\ref{eq49}) and (\ref{eq51}) will then give us
\begin{equation}
p(\omega, \bx, t) 
= \frac{2\nu }{ \omega \Omega (z_{1}^{2} + e^{{-4 \Omega t}}z_{2}^{2})},
\label{eq52}
\end{equation}
for $\alpha<\frac{2 \Omega}{\nu}e^{-2 \Omega t}$. In Eq. (\ref{eq52}), we used
$p(\alpha) = \frac{\nu e^{2 \Omega t}}{2 \Omega}$ for $\alpha= [0, \frac{2 \Omega}{\nu}e^{-2 \Omega t}]$. 
A rapid decrease of $p(\omega, \bx, t)$ in Eq. (\ref{eq52}) for large $z_1^2$ is similar to the elongation of the vorticity profile along 
$z_2$, observed in Figure 5. We note here that 
the condition on $\alpha<\frac{2 \Omega}{\nu}e^{-2 \Omega t}$ is translated into 
$\omega(\bx,t) > \exp{\left[ - \frac{\Omega G_{1}}{2 \nu} e^{{-2 \Omega t}} \right]} \sim \exp{\left[ - \frac{\Omega }{2 \nu}\left( z_{1}^{2} + e^{{-4 \Omega t}}z_{2}^{2}\right) \right]}.$

\medskip
The case ii) { where $\alpha =[\frac{2 \Omega}{\nu}, \alpha_c]$}, we have
$\omega(\bx,t) \sim \frac{2 \Omega}{\alpha \nu} e^{-\Omega t - G_{2}}$,
where $G_{2} = \frac{\Omega}{2\nu} (z_{1}^{2 } + e^{{-2 \Omega t}} z_{2}^{2})$. Thus,
\begin{equation}
p(\omega,\bx,t) \propto \frac{2 \Omega}{\alpha \nu \omega^{2}}e^{-\Omega t - G_{2}},
\label{eq55}
\end{equation}
{ for $\omega =[ \frac{2 \Omega}{\nu \alpha_c} e^{-\Omega t - G_{2}}, e^{-\Omega t - G_{2}}],$}
becoming very small for large $z_1^2$.
Compared with Eq. (\ref{eq50}) or (\ref{eq52}), $p(\omega,\bx,t) \propto \omega^{-2}$ in Eq. (\ref{eq55}) drops more rapidly for large $\omega$. Interestingly, this is similar to the narrowing of Gaussian PDFs
by shear flows shown in \S 4.
Finally, going back to our discussion on the PDF method in \S 2, we can compute the first three terms in
Eq. (\ref{eq7}) using our $p(\omega,\bx,t)$ above to realize that a correct form of the last term in Eq. (\ref{eq7}) is quite complicated and nonlinear in $p(\omega, \bx,t)$,
as noted in \S 2. The diffusion term in Eq.~(\ref{eq3}) cannot be simply neglected and needs to be treated very carefully. 

\section{Discussion and conclusions}
We have presented the first analytical study of the effects of shear flows on enhanced dissipation in a decaying turbulence
{ in 2D}
by incorporating the effects of shear flows non-perturbatively. We considered different initial
power spectra and shear flows (ZF, ZF+ST) and clearly demonstrated how shear flows induce the rapid formation of small scales (large wave number modes), significantly enhancing the dissipation of turbulence. We presented time-dependent PDFs and discussed the effects of enhanced dissipation by shear flows on PDFs and effective dissipation time scale $\tau_{e}$. While previous
works advocated a hybrid time scale $\tau_{e} \propto \Omega^{{-2/3}} \tau_{\eta}$ (e.g. \cite{Diamond}), where $\tau_{\eta}$ is the
time scale due to a molecular diffusion, we showed the dependence of $\tau_{e}$ on $\Omega$ ($\Omega_{z}$) varies with initial power spectra
and also types of shear flows. In addition, we demonstrated the utility of a dynamical time scale $\tau(t)$ in understanding the effect of
shears, which quantifies the rate of the change in information (the rate at which a system passes through statistically different states). 

Overall, $\tau_e$ and $\tau(t)$ tend to be much smaller for an initial $\delta$-function power spectrum and for hyperbolic ZF+ST.  
ZF can dramatically reduce $\tau(t)$ for an initial $\delta$-function power spectrum but not for a constant power spectrum. 
This was however obtained in the case where the mean vorticity $\langle \omega_{T}\rangle$ is independent of time \cite{comment}. A time-varying $\langle \omega_{T}\rangle = -\Omega_{z}$, which is more likely in real situations (e.g. time-varying zonal flows), would however make $\tau(t)$ very small (see Appendix C), with interesting consequences
to be investigated.
Finally, hyperbolic ZF+ ST was shown to cause an exponential increase in wavenumber, with a double exponential decrease in $\langle \omega^{2} \rangle$. 

{ 
The preferential dissipation by shear flows in a certain direction can lead to a strongly anisotropic turbulence, as also shown in \cite{KIM3,KIM5,KIM4}
(with a possibility of the reduction in dimension),
in analogy to the maintenance of a 2D flow in a forced 3D rotating turbulence \cite{Gallet}. 
In 3D, the vortex stretching (which is absent in 2D) could somewhat compensate the severe quenching of vorticity amplitude. 
However, for a linear shear flow
$\U = -x \Omega {\hat y}$, Eq. (\ref{k13}) in Appendix D (see also \cite{KIM5}) shows that the Fourier components of the velocity damp in time as
$\tv_{x} \propto t^{{-2}}e^{-\nu Q(t,0)}$,
$\tv_{z} \propto e^{-\nu Q(t,0)}$, and $\tv_{y} \propto e^{-\nu Q(t,0)}$ to leading order for $t>\frac{k_x(0)}{k_y \Omega}$. Here,
$Q(t,0)= \frac{1}{3} (k_{y} \Omega)^{2} t^{3}+ k_y k_{x }(0)\Omega t^{2}  + [k_x(0)^2+k_y^{2} + k_{z}^{2}]t.$
Therefore, in addition to the
enhanced dissipation $e^{-\nu Q(t,0)}$ through the time-dependent wave number, $v_x$
undergoes the additional algebraic ($\propto t^{-2}$) quenching. The vorticity fluctuation ${\tilde {\bf \omega}}$ would then be at most
$\propto t e^{-\nu Q(t,0)}$ in $y$ and $z$ directions. Investigation of the effect of different shear flows on 3D turbulence, the extension 
to different models such as interchange turbulence \cite{MK17}, magnetic dissipation, and dynamos, and 
implications for extreme events \cite{nature} are left for future work. 
}

\appendix
\section{Relation between ${\cal L}$ and relative entropy}
We first show the relation between $\tau(t)$ in Eq.~(\ref{eq01}) and the
second derivative of the relative entropy (or Kullback-Leibler divergence)
$D(p_{1},p_{2}) = \int dx \,p_{2} \ln{(p_{2}/p_{1})}$ where $p_{1}=p(x,t_{1})$
and $p_{2}=p(x,t_{2})$ as follows:
\begin{eqnarray}
\frac{\partial}{\partial t_{1}} D(p_{1},p_{2}) &=&
 - \int dx p_{2} \frac{ \partial_{t_{1}} p_{1}}{p_{1}},
\label{b1} \\
\frac{\partial^{2}}{\partial t_{1}^{2}}D(p_{1},p_{2}) &=&
 \int dx p_{2} \left[ \frac{ (\partial_{t_{1}} p_{1})^{2}}{p_{1}^{2} }
 - \frac{\partial_{t_{1}}^{2} p_{1}}{p_{1}} \right],
\label{b2}\\
\frac{\partial}{\partial t_{2}} D(p_{1},p_{2}) &=&
 \int dx \left[ \partial_{t_{2} }p_{2}
 + \partial_{t_{2}} p_{2} (\ln{p_{2}}-\ln{p_{1}}) \right ],
\label{b3} \\
\frac{\partial^{2}}{\partial t_{2}^{2}} D(p_{1},p_{2}) &=&
 \int dx \left[ \partial_{t_{2}}^{2} p_{2} 
 + \frac{(\partial_{t_{2}} p_{2})^{2}}{p_{2}}
 + \partial_{t_{2}}^{2} p_{2} (\ln{p_{2}}-\ln{p_{1}}) \right ].
\label{b4} 
\end{eqnarray}
By taking the limit where $t_{2} \to t_{1} = t$ ($p_{2} \to p_{1}=p$) and by using
the total probability conservation (e.g. $\int dx \partial_{t } p = 0$),
Eqs.~(\ref{a1}) and (\ref{a3}) above lead to
$$ \lim_{t_{2} \to t_{1}=t} \frac{\partial}{\partial t_{1}} D(p_{1},p_{2})
 =\lim_{t_{2} \to t_{1}=t} \frac{\partial}{\partial t_{2}} D(p_{1},p_{2})
 = \int dx \partial_{t } p=0,$$
while Eqs.~(\ref{a2}) and (\ref{a4}) give
$$ \lim_{t_{2} \to t_{1}=t} \frac{\partial^{2}}{\partial t_{1}^{2}}D(p_{1},p_{2}) 
= \lim_{t_{2} \to t_{1}=t} \frac{\partial^{2}}{\partial t_{2}^{2}} D(p_{1},p_{2})
= \int dx \frac{ (\partial_{t} p)^{2}}{p }.$$

To link this to information length ${\cal L}$, we then express $D(p_{1},p_{2})$
for small $dt = t_2 - t_1$ as
\begin{equation}
D(p_{1},p_{2}) = \left[ \int dx
 \frac{ (\partial_{t_1} p(x,t_1))^{2}}{p } \right] (dt)^2 + O((dt)^3),
\label{b5}
\end{equation}
where $O((dt)^3)$ is higher order term in $dt$. We define the infinitesimal
distance (information length) $dl(t_1)$ between $t_1$ and $t_1 +dt$ by
\begin{equation}
dl(t_1) = \sqrt{ D(p_{1},p_{2})} = \sqrt{ \int dx
 \frac{ (\partial_{t} p)^{2}}{p }} dt + O((dt)^{3/2}).
\label{b6}
\end{equation}
The total change in information between time $0$ and $t$ is then obtained by
summing over $dt(t_1)$ and then taking the limit of $dt \to 0$ as
\begin{eqnarray}
{\cal L}(t) &= &\lim_{dt \to 0} \left [ dl(0) +dl(dt) +dl(2dt) + dl(3dt)
 + \cdot\cdot \cdot dl(t-dt) \right]
\nonumber \\
& = & 
\lim_{dt \to 0} \left [ \sqrt{ D(p(x,0),p(x,dt))} + \sqrt{ D(p(x,dt),p(x,2dt))}
 + \cdot\cdot\cdot \sqrt{ D(p(x,t-dt),p(x,t))} \right ]
\nonumber \\
& \propto& \int_0^t dt_1\, \sqrt{ \int dx \frac{ (\partial_{t_1} p)^{2}}{p }}.
\label{ab}
\end{eqnarray}

\section{Anisotropic constant power spectrum}
To demonstrate an incoherent shearing effect in the presence of multiple modes, it is interesting to consider an isotropic power spectrum by keeping a constant spectrum in $k_{y}$ but taking $k_x(0) \sim 0$. The mean square vorticity is obtained from Eq. (\ref{eq24}) by taking $k_x(0) \to 0$, with the result
\begin{eqnarray}
\left \langle \omega^{2}(x,t)\right \rangle
&=& \sqrt{\frac{\pi}{2 \nu t (1 + \frac{1}{3} \Omega_{z}^{2} t^{2})}}\phi.
\label{eq240}
\end{eqnarray}
Thus, $\left \langle \omega^{2}(x,t)\right \rangle \propto t^{-3/2}$, decreasing less rapidly than $\left \langle \omega^{2}(x,t)\right \rangle \propto t^{-2}$ in Eq. (\ref{eq24}). On the other hand, the effective dissipation time $\tau_{e}$ is similar to Eq. (\ref{eq25}).

\section{Slowly time-varying ZF}
We assume $\Omega_{z} = \Omega_{z0}e^{-t/\tau_{0}}$ and $k_{x0} \sim 0$. Then, we have
\begin{eqnarray}
k_{x}(t) &=& \int_{0}^{t}dt_{1}k_{y} \Omega_{z}(t_{1}) = k_{y}\Omega_{z0}\tau_{0}(1-e^{{-t/\tau_{0}}}),
\label{eq500}\\
Q_{1}(t) &=& (k_{y}\Omega_{z0}\tau_{0})^{2}\frac{\tau_{0}}{3} \left[1-e^{-t/\tau_{0}}\right]^{3} + k_{y}^{2}t
\sim \frac{1}{3}  (k_{y}\Omega_{z0})^{2} t^{3} + k_{y}^{2}t,
\label{eq501}\\
\partial_{t} \Omega_{z} &\sim & -\frac{1}{\tau_{0}} \Omega_{z0},
\label{eq502}
\end{eqnarray}
for $t \ll \tau_{0}$.  Thus, Eqs. (\ref{eq003}), (\ref{eq24}) and (\ref{eq250}) with the help of Eqs. (\ref{eq501})-(\ref{eq502}) give us
\begin{eqnarray}
{\cal E} = \frac{1}{\tau(t)^{2}} =  \frac{1}{2} \frac{(\p_{t}\beta)^{2}}{\beta^{2}} 
+ 2 \beta (\p_{t} \Omega_{z})^{2 } 
&\sim & \frac{(4 + \frac{2}{3} \Omega_{z}t^{2} )^{2} }{2 t^{2}( 4 +\frac{1}{3} \Omega_{z}t^{2} )^{2}}
+  \frac{2 \nu t \sqrt{4 + \frac{1}{3} \Omega_{z0}t^{2}}}{\pi \phi \tau_{0}^{2}} \Omega_{z0}^{2}.
\label{eq503}
\end{eqnarray} 
The second term is due to the change of $\Omega_{z}$ measured
in the unit of the very small PDF width $\propto \beta^{-\frac{1}{2}} \propto \langle \omega^{2} \rangle^{\frac{1}{2}}$. 
As time increases, the second term obviously makes a significant contribution.

{
\section{3D hydrodynamic turbulence \cite{KIM5}}
In 3D, the main governing equations for the total velocity ${\bf u}={\bf v} +\U$ are 
\begin{eqnarray}
\p_t \bu + \bu \cdot \nabla \bu 
&=& -\nabla p + \nu \nabla^2 \bu + {\bf f}\,,
\label{k1}\\
\nabla \cdot \bu &=&  0\,,
\label{k2}
\end{eqnarray}
where ${\bf f}$ is a small scale forcing in general.  By using $\U = -x \Omega {\hat y}$
\begin{eqnarray}
\p_t \hv_x &=& -i k_x {\hat p}    + \hf_x \,,
\label{a1}\\
\p_t \hv_y -\Omega \hv_x &=& -i k_y {\hat p}  + \hf_y\,,
\label{a2}\\
\p_t \hv_z  &=& -i k_z {\hat p}  + \hf_z\,,
\label{a3}\\
0&=& k_x \hv_x + k_y \hv_y + k_z \hv_z \,,
\label{a4}
\end{eqnarray}
where the second term in Eq. (\ref{a2}) is due to the vortex stretching.
Here, ${\hat w}$ and ${\tilde w}$ for $w=v_i,p$ and $f$ are defined as
\begin{eqnarray}
{w} ({\bf x},t) && = \tilde{w}({\bf k},t) \exp{\{i(k_x(t) x + k_y y + k_z z)\}}\,,
\label{a6}\\
{\hat w}&& \equiv {\tilde w} \exp{\{\nu (k_x^3/3 k_y \Omega + k_H^2 t)\}}\,,
\label{a9}
\end{eqnarray}
where $k_H^2 = k_y^2 + k_z^2$; $k_x(t)=k_{x}(0)+\Omega k_{y}t$.
Now, to solve coupled equations (\ref{a1})--(\ref{a4}), we introduce
a new time variable $\tau = k_x/k_y + \Omega t$ and rewrite
them as:
\begin{eqnarray}
\Omega \p_\tau \hv_x &=& -i \tau k_y {\hat p}    + \hf_x ,
\label{a7}\\
\Omega \p_\tau \hv_y -\Omega \hv_x &=& -i k_y {\hat p}  + \hf_y,
\label{a8}\\
\Omega \p_\tau \hv_z  &=& -i k_z {\hat p}  + \hf_z,
\label{a9}\\
0&=& \tau \hv_x + k_y \hv_y + {k_z \over k_y} \hv_z.
\label{a10}
\end{eqnarray}
A straightforward, but rather long, algebra then gives us the solutions in
the following form: 
\begin{eqnarray}
\hv_x(\tau) &=&
{1\over \gamma + \tau^2}
\int ^\tau d\tau_1 h_1(\tau_1),
\nonumber \\
\hv_z(\tau) &=&
\int ^\tau d\tau_1 
\left [ {\bbeta\over \tau_1} \hv_x - {\bbeta \over \tau_1} \hf_x + \hf_z
\right],
\nonumber\\
&=& -{\bbeta \tau \over \gamma} \hv_x
+ \int ^\tau d\tau_1 
{1 \over \gamma} \left [ h_2(\tau_1) -{\bbeta \over \gamma^{1/2}} 
\left(\tan^{-1}{{\tau\over \sqrt{\gamma}}} -
\tan^{-1}{{\tau_1\over \sqrt{\gamma}}}\right) h_1(\tau_1)\right],
\nonumber \\
{\hv}_y(\tau) &=&
-\tau {\hv}_x(\tau) 
-\bbeta u \hv_z(\tau) ,
\nonumber \\
{\hat p} &=& {\Omega \over k_y} (-\p_\tau \hv_x + \hf_x),
\label{a14}
\end{eqnarray}
where $\bbeta=k_z/k_y$, $\gamma = 1 + \bbeta^2$, 
$h_1=  (1+\bbeta^2)\hf_x - \tau \hf_y - \tau \bbeta \hf_z$, and
$h_2=  - \bbeta \hf_y + \hf_z$.
Finally, going back to the original variable $k_x = k_y \tau$,
we obtain
\begin{eqnarray}
\tv_x(\bk(t),t)
&=& \int dt_1 d^3 k_1 
{k_y^2 \over k^2} 
{\hat g}(\bk,t;\bk_1,t_1) e^{-\nu Q(t,t_1)}
{\tilde h}_1(\bk_1,x, t_1),
\nonumber \\
\tv_z(\bk(t),t)
&=& -{k_x k_z \over k_H^2} \tv_x(\bk(t),t)
+ \int dt_1 d^3 k_1 
{\hat g}(\bk,t;\bk_1,t_1) e^{-\nu Q(t,t_1)}
{\tilde h}_2(\bk_1,x, t_1)
\nonumber \\
&&
\times \left[ {k_y^2 \over k_H^2}  
{\tilde h}_2(\bk_1,x, t_1)
- {k_z k_y^2 \over |k_H^3|} \left[ \tan^{-1}{\left(
{k_x\over |k_H|}\right)}
-\tan^{-1}{\left({k_{1x}\over |k_{1H}|}\right)} \right]
{\tilde h}_1(\bk_1,x, t_1)
\right],
\nonumber \\
\tv_y(\bk(t),t)
&=& -{k_x \over k_y} \tv_x(\bk(t),t)
-{k_z \over k_y} \tv_z(\bk(t),t)
\,.
\label{k13} 
\end{eqnarray}
Here, $Q(t,t_1) =  \int^t_{t_1} dt' [k_x^2(t') + k_H^2] =
[k_x^3-k_{1x}^3] /3k_y\Omega + k_H^2 (t-t_1)$; 
$k_H^2 = k_y^2 + k_z^2$; $k^2=k_H^2+ k_x^2$;  
${\hat g}(\bk,t;\bk_1,t_1)
= \delta(k_y-k_{1y}) 
\delta (k_z-k_{1z}) 
\delta
\left[k_{x} -k_{1x} - k_{1y} (t-t_1) \Omega\right]$;
$\hh_1 = (1+k_z^2/k_y^2) \tf_x - k_x \tf_y/k_y - k_x k_z \tf_z/k_y^2$;
$\hh_2 = - k_z \tf_y/k_y + \tf_z$.
By taking $\tf_{i} (t_{1})= \tv_{i} (t_{1}) \delta(t_{1})$, we obtain the homogeneous solution without the forcing.
}

\end{document}